# A drop model of integer and fractional quantum Hall effects

A.A. Vasilchenko

*National Research Tomsk State University, 634050 Tomsk, Russia*

Abstract

We present a drop model for integer and fractional quantum Hall effects (FQHE). We show that the two-dimensional electron gas breaks up into regions with filling factors $v = 1$ and $v = 0$ in disk geometry, and the formation of drops with a finite number of electrons is possible. Sequences of filling fractions are constructed on the basis of experimental data. For all sequences there are initial FQHE states, which correspond to a drop with five electrons. The remaining FQHE states are composite states of a drop with five electrons and one or more pairs of electrons.

**1. Introduction**

Existing models of the fractional quantum Hall effect (FQHE) mainly explain states at fractional filling factors with odd denominators [1, 2]. It is well known that FQHE is observed at even denominators as well [3-7]. Before the discovery of the quantum Hall effect, it was shown that a two-dimensional electron gas in a magnetic field at a fill factor $v < 1$ breaks up into spatially inhomogeneous regions [8]. Such partitioning is also possible at the filling factor $v = 1$ [9].

In this work we propose a new model in which the integer quantum Hall effect (IQHE) and FQHE are explained from unified positions. We believe that a two-dimensional electron gas in a magnetic field at $v \leq 1$ breaks up into regions with $v = 1$ and $v = 0$. In this case the Coulomb interaction increases energy and the exchange interaction lowers energy and drops with a finite number of electrons $N$ should be formed.

For a drop with a number of electrons of the order of ten or less, the transverse dimensions will be comparable to the width of the quantum well, so calculations must be performed for a three-dimensional drop. In addition, the results of the experiment [10] indicate that the electron density has a quasi-one-dimensional character in two-dimensional quantum dots.

In this work we consider a simplified model and the results are qualitative in nature. We assume that the two-dimensional electron layer has zero thickness and the drops have circular symmetry in the *xy*-plane. We apply the density-functional theory (DFT) to determine the ground states of a two-dimensional drop in the presence of an external magnetic field.

## 2. Theoretical model

We use effective atomic units. Energy is expressed in units of $Ry = e^2/(2\varepsilon a_B)$, and length in units of $a_B = \varepsilon\hbar^2/(m_e e^2)$, where $m_e$ is the effective electron mass, $\varepsilon$ is the dielectric constant. In the calculations, we have used the material constants of GaAs, for which $\varepsilon = 12.4$ and $m_e = 0.067 m_0$ ($m_0$ is the free electron mass). For GaAs we get $a_B = 9.8$ nm, $Ry = 5.9$ meV.

We consider $N$ two-dimensional electrons confined by an external potential created by a uniform positively charged background with density $n_p$. The drop radius is found from the expression $n_p \pi R^2 = N$. The total energy for such a system is

$$E[n] = T[n] + E_{ext}[n] + E_H[n] + E_{xc}[n], \tag{1}$$

where $T[n]$ is kinetic energy of non-interacting electrons in magnetic field B, which is given by vector potential $\mathbf{A} = B(-y/2, x/2)$.

The second term in expression (1) is given by

$$E_{ext}[n] = \int V_{ext}(r) n(r) d\mathbf{r}, \tag{2}$$

where $V_{ext}(r) = 2\int_0^R \frac{n_p}{|\mathbf{r}-\mathbf{r}'|} d\mathbf{r}'. \tag{3}$

The Coulomb energy is

$$E_H[n] = \frac{1}{2}\int V_H(r) n(r) d\mathbf{r}, \tag{4}$$

where $V_H(r) = 2\int_0^\infty \frac{n(r')}{|\mathbf{r}-\mathbf{r}'|} d\mathbf{r}', \tag{5}$

For the exchange energy we use the local-density approximation and exclude the self-interaction of electrons:

$$E_x[n] = \int \varepsilon_x(n) d\mathbf{r} - \sum_m \int \left(\varepsilon_x(n_m) + \frac{1}{2}V_{H,m}(r)\right) n_m(r) d\mathbf{r}, \tag{6}$$

where $V_{H,m}(r) = 2\int_0^\infty \frac{n_m(r')}{|\mathbf{r}-\mathbf{r'}|}dr'$, $n_m(r)$ is the density of the $m$-th electron, $\varepsilon_x(n)$ is the exchange energy per electron for a homogeneous electron gas, which for the lower spin Landau level has the following form:

$$\varepsilon_x(n) = -\sqrt{2\pi}\pi L n(r), \qquad (7)$$

here $L$ is the magnetic length.

By minimizing functional (1), one obtains the Kohn-Sham equations for spin-polarized electrons

$$\{-\frac{\partial^2}{\partial r^2} - \frac{1}{r}\frac{\partial}{\partial r} + \frac{r^2}{4L^4} + \frac{m^2}{r^2} - \frac{m}{L^2} + V_{eff}(r)\}\psi_m(r) = E_m\psi_m(r), \qquad (8)$$

with the effective single-particle potential

$$V_{eff}(r) = V_H(r) - V_{H,m}(r) + 2\alpha(n(r) - n_m(r)) + V_{ext}(r), \qquad (9)$$

where $m$ is the angular momentum of electron, $n_m(r) = |\psi_m(r)|^2$, $n(r) = \sum_{occ\, m} n_m(r)$, $\alpha = -\sqrt{2\pi}\pi L$.

## 3. Numerical results and discussion

The Kohn-Sham equations (8) – (9) are solved numerically. For comparison with the results of exact diagonalization [11, 12], we performed calculations with the confining potential $V_{ext}(r) = \omega_0^2 r^2/4$. For quantum dots with $N = 3$ [11] and $N = 7$ [12], the differences in energy were three and thirteen percent, respectively, and the same magic numbers for the total angular momentum of the electrons were obtained. Note that in all cases the transition to the new angular momentum state in DFT takes place in higher magnetic fields compared to the results of exact diagonalization [11-18].

We perform calculations for $n_p = 10^{11}$ cm$^{-2}$. Figures 1, 2, 3, and 4 show the total energy as a function of the total angular momentum $M$ for $N = 5, 8, 10$, and 13. In all cases the magic angular momentum is given by

$$M = M_0 + p(N-k), \qquad (10)$$

where $M_0 = N(N-1)/2$; $p = 0, 1, 2, ...$; $k = 0, 1, 2, ...$.

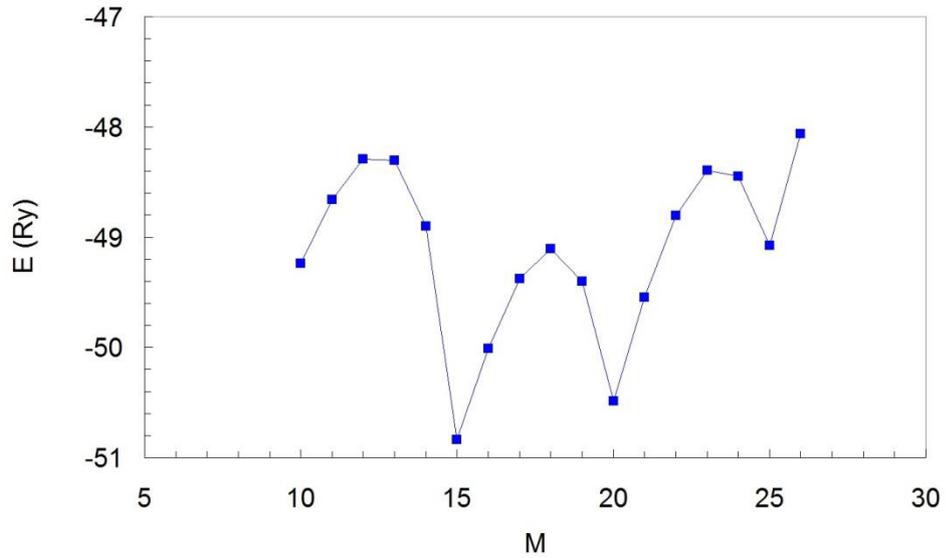

Figure 1. Total energy as a function of the total angular momentum; $N = 5$, $B = 10$ T. Lines are guides to the eye.

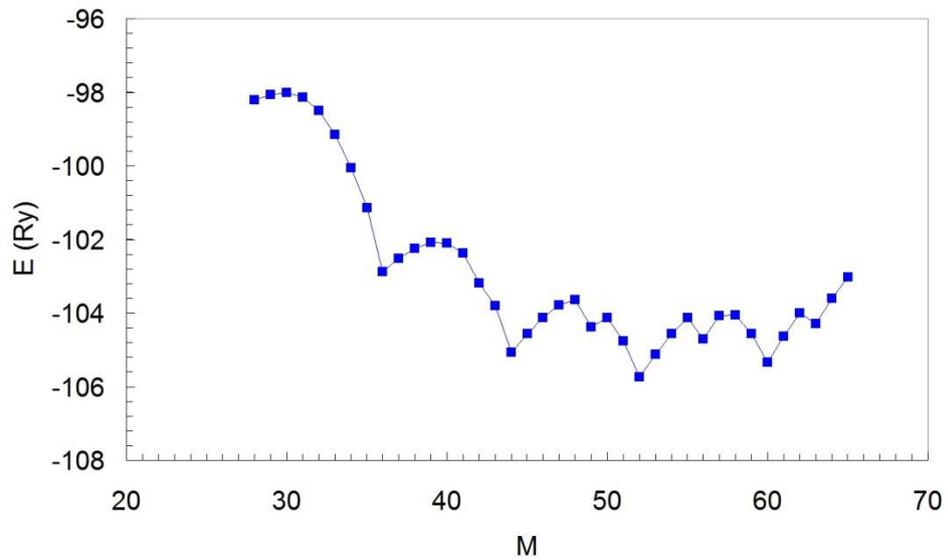

Figure 2. Total energy as a function of the total angular momentum; $N = 8$, $B = 10$ T. Lines are guides to the eye.

For that set of magic numbers, $k$ electrons are in the center of the drop, and the remaining electrons are in the ring and have a compact configuration. For $N = 5$ electrons all magic numbers belong to the state with $k = 0$, and all electrons have a compact configuration. In the one-particle approximation the transition to a new state occurs when the electron with the minimal angular momentum transitions to an unfilled outer level. For $N = 8$ electrons, new magic numbers 49, 56 and 63 appear, which refer to the state with $k = 1$. In this state one electron is in the center of the drop, and the others are distributed throughout the ring. Note that these states are metastable. The ground state with $k = 1$ will be the state with $M = 70$ at $B = 12.6$ T. As $N$ decreases, the transition from the $k = 0$ state to the $k = 1$ state occurs in higher magnetic fields. For $N = 6$ this transition occurs at $B = 18.6$ T, and for $N = 5$ at $B = 22.9$ T.

For $N = 10$ electrons (Figure 3), we have magic angular moments with $k = 0, 1$ and 2. The state with $k = 1$ will become ground at $B = 8.9$ T. The states with $k = 2$ ($M = 69, 77, 93$ and 101) remain metastable at any magnetic fields. The ground state with $k = 2$ will be the state with $M = 117$ at $B = 12.2$ T. Metastable states play an important role because they lead to a decrease in the value of the energy gap (the energy difference between the ground state and the first excited state).

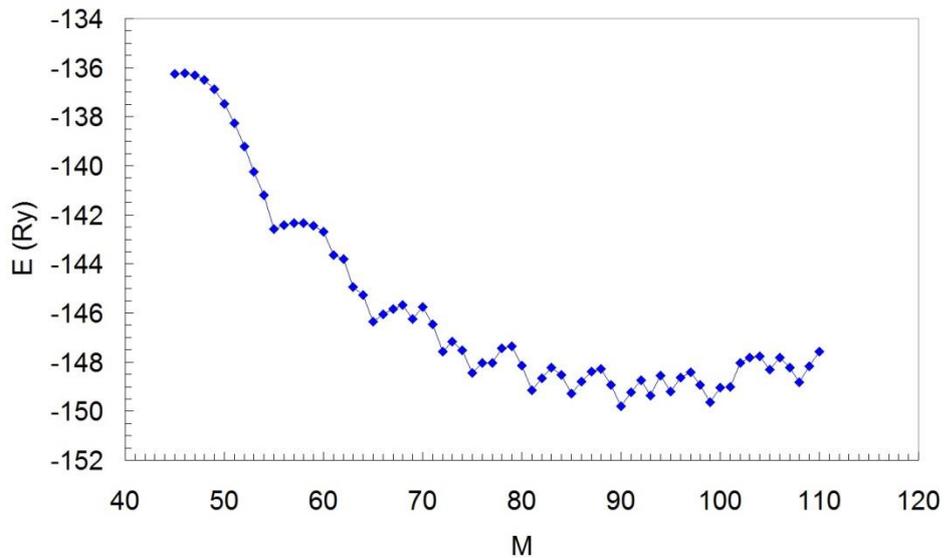

Figure 3. Total energy as a function of the total angular momentum; N = 10, B = 10 T. Lines are guides to the eye.

As $N$ increases, new sets of magic numbers with $k > 2$ appear, and the value of the energy gap decreases. At $N = 13$ (Figure 4) there are sets of magic numbers with $k = 0, 1, 2$ and 3. Most of these states will remain metastable. Ground states at $B < 10$ T can be only states with $k = 0$ and $k = 2$. Note that transition to states with $k = 2$ takes place at rather low magnetic field $B = 6.7$ T. New magic numbers 141, 171, 181, 190, and 194 also appear. At these magic numbers, states with $m = 0$ are not occupied, but states with $m = 1, 2, 3$ are occupied. The configuration of electrons is (1,2,3,9,10,11,12,13,14,15,16,17,18) for $M = 141$, (1,2,12,13,14,15,16,17,18,19,20,21,22) for $M = 190$, and (2,3,4,14,15,16,17,18,19,20,21,22,23) for $M = 194$. For these series of magic numbers

$$M = M_0 + p(N-k) + rN, \tag{11}$$

where $r = 0, 1, 2, ....$

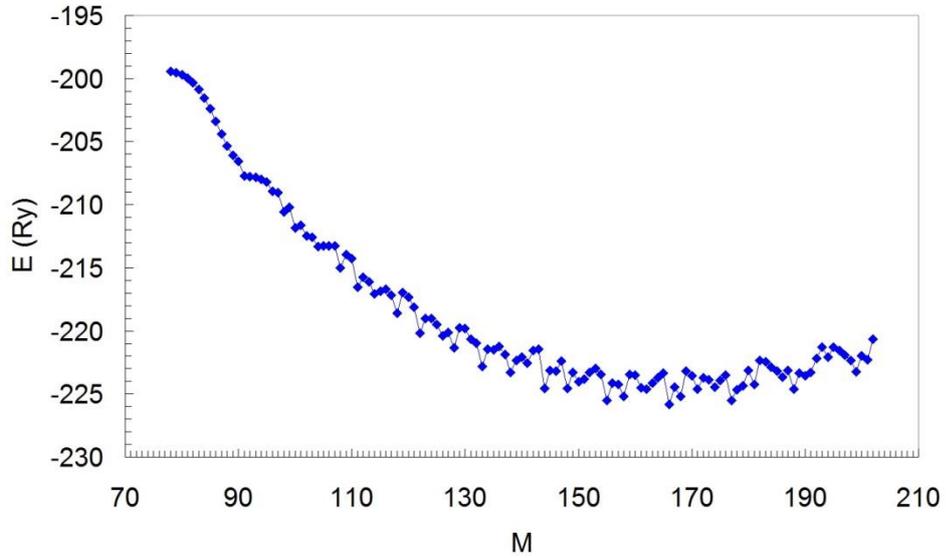

Figure 4. Total energy as a function of the total angular momentum; $N = 13$, $B = 10$ T. Lines are guides to the eye.

As the number of electrons in the drop increases, the electrons will be distributed into rings, and they have a compact configuration inside each ring. In this case we can assume that the two-dimensional electron gas breaks up into regions with $v = 1$ and $v = 0$. This result is in qualitative agreement with the results of [8]. Inside each ring with $v = 1$, the exchange interaction tends to increase the electron density along the ring and a symmetry breaking with the formation of a charge density wave or electron drops is possible.

Calculation results for the disk geometry show that as the number of electrons in the drop increases, the energy per electron decreases. Thus, the energy difference per electron for drops with $N = 5$ and $N = 6$ is equal to 1.1 meV. As $N$ goes up this difference decreases and for drops with $N = 13$ and $N = 14$ it is equal to 0.7 meV. Results of calculations for disk geometry show that the kinetic energy per electron almost does not change with changes in the number of electrons in the drop and the main contribution to energy change is made by Coulomb energy. The greatest contribution to the energy change is made by the Coulomb energy of interaction of electrons with the background charge. Therefore, the shape of the drop must be determined from the minimum of Coulomb energy. For a quasi-one-dimensional drop, the contribution of Coulomb interaction will be less [19]. Experimental data [10] also testify to the quasi-one-dimensional character of the electron density in quantum dots.

Let us define the fill factor as [20]

$$\nu = \frac{M_0}{M} \qquad (12)$$

For states with $k = 0$ the filling factor $\nu = 1/q$ with odd denominator $q$ is possible at any values of $N$, while states with even denominator $q$ are possible only at odd values of $N$. Figure 5 shows the dependence of filling factor on the magnetic field for the number of electrons $N = 5, 7$ and $9$. At $M = M_0$ a IQHE plateau is observed. With increasing magnetic field there is a transition to the FQHE state with filling factor $\nu = M_0/M$. Calculations have shown that the width of the plateau decreases in the region of transition from state $k = 0$ to state $k = 1$. For this reason, the plateaus $\nu = 1/3$ for $N = 7$ and $\nu = 1/2$ for $N = 9$ are narrow. As indicated earlier, the results obtained using DFT are not consistent with the results of exact diagonalization. In DFT, the transition to a new state occurs at higher magnetic fields and, most importantly, the filling factor (12) differs from the value of the filling factor for the macroscopic system $\nu = n_p 2\pi L^2$. Therefore, the results presented in Figure 5 are qualitative in nature. Results of the exact diagonalization and FQHE experiments show that the middle of all plateaus has an almost directly proportional dependence on the magnetic field:

$$M_p = cB_p, \qquad (13)$$

where $c$ is a constant.

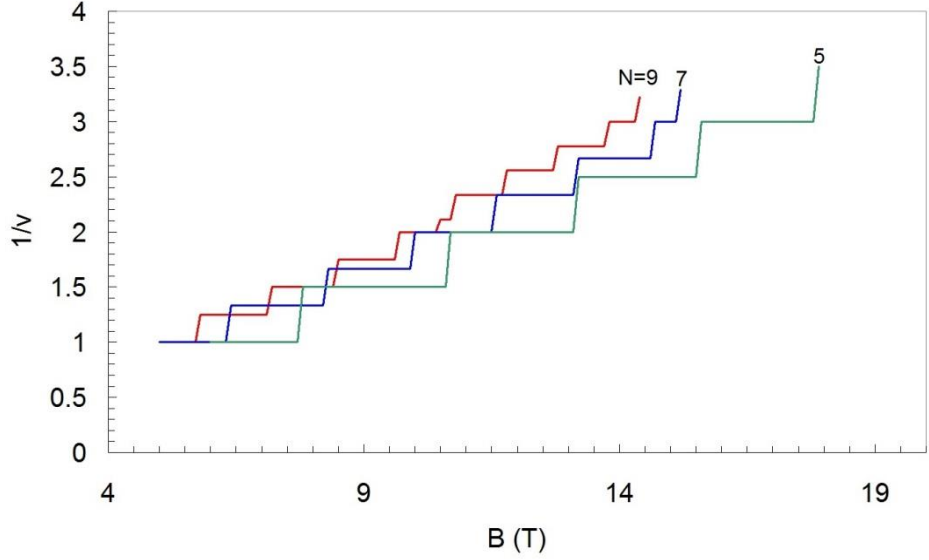

Figure 5. Magnetic field dependence of $1/v = M/M_0$ for $N = 5$, 7 and 9 electrons.

Taking into account that $v=n_p 2\pi L^2$ and using (13) we obtain expression (12). From expressions (10) and (13) we obtain for the plateau width at $k = 0$

$$\Delta B = \frac{2B_0}{N-1} . \qquad (14)$$

Note that the plateau widths in Figure 5 also decrease with increasing $N$ and are close to dependence (14).

From expression (14) we obtain for the width of the plateau by magnetic flux $\Delta \Phi = \Delta B \pi R^2$:

$$\Delta \Phi = \frac{2N}{N-1}\frac{h}{e}. \qquad (15)$$

The change in the angular momentum of the electrons leads to the oscillations of the persistent current. The oscillation period of the persistent current $\Delta \Phi$ can vary from four to two quanta of magnetic flux. It is of interest to study experimentally the persistent current oscillations in the quantum Hall effect state.

The energy gap $\Delta$ between the ground state and the first excited state oscillates as a function of the magnetic field and has its maximum in the middle of the plateau. The transition from the $v = 1$ state to the unpolarized state has not been studied, and we assumed the width of the $v = 1$ plateau to be equal to the width of the $v = (N-1)/(N+1)$ plateau. The energy gap in the center of the plateau at $v = 1$ for a drop with $N = 5$ is 1.5 meV and changes insignificantly with increasing number of electrons ($\Delta = 1.3$ meV for $N = 13$). The energy gap decreases during the transition to FQHE state and is equal to 0.8 meV at $v = 2/3$ for $N = 5$. As the magnetic field increases, the maximum energy gap decreases and is equal to 0.5 meV at $B = 11.9$. At $N < 10$ the value of the maximum energy gap weakly depends on the number of electrons in the drop. The presence of metastable states and the transition to the states with new $k$ lead to a decrease in the energy gap. Thus, at $N = 6$ and $M = 27$ the maximum energy gap is equal to 0.6 meV, whereas at $N = 13$ and $M = 166$ the energy gap is two times smaller.

More than fifty FQHE states have been discovered for $v < 1$ [6]. Most of them correspond only to the minimum of the longitudinal resistance. In drop model, FQHE states with the widest plates $v = 2/3, 3/5, 3/7, 2/5, 1/3, 2/7$ occur in drops with a number of electrons $N = 4$ and 5. Drops with $N < 10$ account for most of the FQHE states. Drops with $N > 9$ correspond to FQHE states near $v = 3/4$ and $v = 1/q$ ($q = 2, 3, 4$). Analysis of the experimental data [6] shows that there are sequences of fractions $(3l+1)/(4l+1)$ and $(3l-1)/(4l-1)$ for states $v = 3/4$ ($l$ is a positive integer) and $l/(ql\pm1)$ for states $v = 1/q$ ($l \geq 2$). Most FQHE states are located near $v = 1/2$: 2/3, 3/5, 4/7, 5/9, 6/11, 7/13, 8/15, 9/17, 10/19, 10/21, 9/19, 8/17, 7/15, 6/13, 5/11, 4/9, 3/7, 2/5. As we approach $v = 1/2$, the number of electrons in the drop should increase, with the number of electrons in the drop changing by two electrons in the transition to the new state. As the quality of samples improves and the temperature goes down, new FQHE states should appear. For example, for a sequence with $v = 1/2$, new FQHE states 11/21, 11/23, etc. will appear. The sequence for $v = 1/3$ is as follows [6]: 2/5, 3/8, 4/11, 6/17, 4/13, 3/10, 2/7. Note that these states, except for 6/17, belong to drops with a number of electrons $N < 10$, while state 5/14 is not observed in the experiment. It is unclear why FQHE states $1/q$ with an even denominator do not exist. A FQHE state with odd denominator $q$ is possible in drops with any number of electrons, while for even denominator $q$ this state is formed only at odd values of *N*. Perhaps the FQHE state with an odd denominator *q* is formed in drops with even value of *N* and in this case Bose condensation of electrons occurs. Another more realistic option is also possible. The presence of impurities, disorders, inhomogeneity of positively charged background and boundaries of the sample may lead to the formation of drops with different numbers of electrons (multicomponent electronic liquid). The transition from the *l* state to the *l*+1 state does not involve the addition of two electrons to the drop, but a composite state of a drop with *N* electrons and a drop with two electrons appears. When approaching the $v = 1/q$ state, at odd denominator *q*, the electron pairs condense into a Bose liquid and a drop with an even number of electrons is formed.

The small number of observable FQHE states in sequences with $q \geq 5$ is due to the fact that the energy gap decreases as the magnetic field increases. There are initial states for each sequence: $v = 2/3$ and 2/5 for $v = 1/2$ sequence; $v = 2/5$ and 2/7 for $v = 1/3$ sequence; $v = 2/7$ and 2/9 for $v = 1/4$ sequence, etc. For each sequence, the "initial" drop contains 5 electrons. All other fractions in the sequences are explained by the composite states of the "initial" drop and one or more pairs of electrons. For example, for $v = 3/8$, a drop with five electrons and a pair of electrons would be a composite state; for $v = 4/11$ the drop has five electrons and two pairs of electrons. One would expect that in IQHE state, at $v = 1$, drops with five electrons would form.

There are only three observable states ($v = 7/11, 5/13, 5/17$) that are not part of the sequences. These FQHE states are near the initial states and are possible at $N = 8$ and $N = 6$. By analogy with these states, FQHE states can also exist: $v = 9/11$, 9/13, 5/8 [5], 7/17, 7/19, 5/23.

Figure 6 shows the electron density profiles in a drop with $N = 5$. For clarity, the graph is plotted up to the boundary of the drop $r = R$. It can be seen that the electron density is very low at the boundary of the drop. Note that at $B = 6.3$ (plateau center at $v = 1$) the electron density at the border of the drop is less than the electron density at the center of the drop by about an order of magnitude. At *B*

= 23.5 there is one electron at the center of the drop and this state corresponds to the $k = 1$ state.

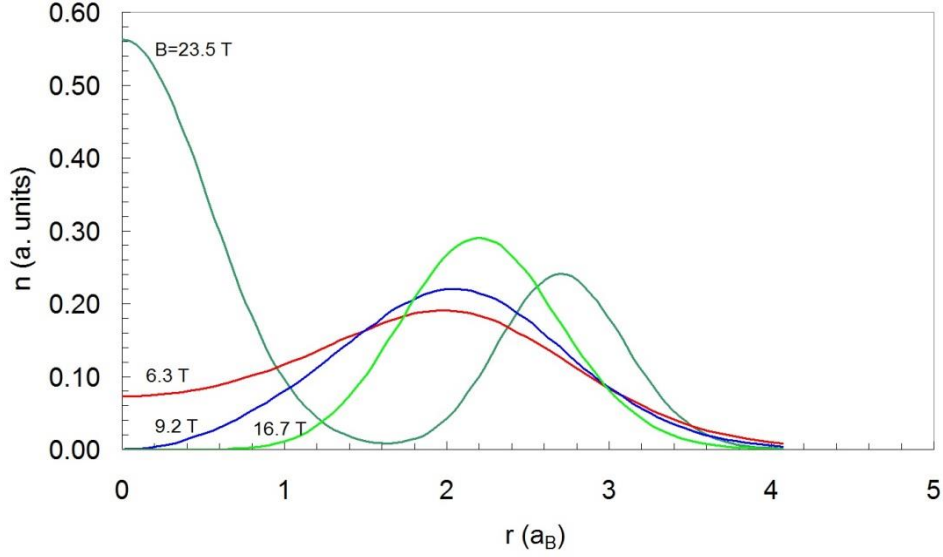

Figure 6. Density profiles of electrons in a drop with $N = 5$ electrons.

The FQHE states at $\nu > 1$ are also explained by the drop model with fully filled lower Landau levels. We need a new parameter to explain the observed states with even denominators ($\nu = 3/2, 5/2, 9/2$). Spin polarization of electrons can be accepted as such a parameter [21]. The spin polarization of electrons has been experimentally confirmed in studies in ZnO quantum wells at $\nu = 3/2$ [4]. The $\nu = 3/2$ state can also be influenced by the narrowing of the electron channel [7]. The narrowing of the electron channel should lead to rearrangements of the energy spectrum and the shape of the electron drop and a state with an even denominator may become possible.

## 4. Conclusions

The ground states of $N$-electron drops have been found to study IQHE and FQHE. It is shown that as the number of electrons in the drop increases, the plateau widths decrease. Energy gaps were calculated and it was found that their values decrease with increasing magnetic field. Sequences of fractions for $\nu = 3/4$ and $\nu = 1/q$ were constructed on the basis of the experimental data. For $\nu = 1/q$ sequences, as filling factor is approaching $\nu = 1/q$, each successive FQHE state contains two more electrons than the previous one. For all sequences there is an initial state, which corresponds to a drop with five electrons. All other FQHE states are composite states of a drop with five electrons and one or more pairs of electrons. For odd denominator $q$ a transition to an even number of electrons in the drop is possible and Bose condensation of electrons occurs. It can be expected that in the IQHE state at $v = 1$, drops with five electrons are also formed. The spatial inhomogeneity of the potential relief may lead to the formation of drops with different numbers of electrons.

The shape of the drop must be determined by a minimum of Coulomb energy. Apparently, the density of electrons in the drop must be quasi- one-dimensional. This is also indicated by experimental results [10]. For quantitative analysis it is necessary to perform calculations in three dimensions for a quasi-one-dimensional drop. One transverse dimension of the electron drop will be determined by the width of the quantum well, and the second transverse dimension will be equal to several magnetic lengths. For a quasi-one-dimensional drop there will be a shell structure of the electron drop, and the transition to a new state occurs when an electron moves from one energy level to another, while the compact configuration of electrons is preserved.

In a zero magnetic field, electron drops can be formed in quasi-two-dimensional layers. It was shown experimentally in [22-25] that a quasi-two-dimensional electron liquid is spatially inhomogeneous. Particularly noteworthy are the results of [24], which show the possibility of forming spin drops with a number of electrons $N = 4$ and total spin $S = 2$. The formation of the electron drops in low-dimensional systems may lead to a new mechanism of superconductivity.

**Acknowledgements**

This work was supported by the State Assignment of the Ministry of Science and Higher Education of the Russian Federation (project No. 0721-2020-0048).

E-mail: a_vas2002@mail.ru